\documentclass[reprint, prmaterials, amsfonts, amssymb, amsmath, aps, floatfix, article]{revtex4-2}
\usepackage[pdfusetitle]{hyperref}
\usepackage{graphicx}% Include figure files
\usepackage{dcolumn}% Align table columns on decimal point
\newcolumntype{d}[1]{D{.}{.}{#1}}
\usepackage{bm}% bold math
\usepackage{longtable}
\usepackage{mathtools}

\usepackage{xspace}
\usepackage{xpunctuate}
\DeclareRobustCommand\etal{\xperiodafter{\emph{et al}}}

\hypersetup{
  colorlinks=true,
  linkcolor=blue,
  filecolor=blue,
  citecolor=blue,
  urlcolor=blue,
}

\newcommand{\lno}{La$_3$Ni$_2$O$_7$\xspace}

\usepackage{float}

\begin{document}

\title{Pressure-tunable structural instabilities in
  single-layer-trilayer La$_3$Ni$_2$O$_7$}

% \title{Nearly-degenerate distinct low-symmetry phases due to the $\mathbf{
%     \textit{q} =(\frac{1}{3},\frac{1}{3},\frac{1}{3})}$ phonon
%   instability in kagome metal ScV$_6$Sn$_6$}

% \title{Nearly-degenerate minima in the four-dimensional order 
%   parameter subspace spanned by the $\mathbf{ \textit{q}
%     =(\frac{1}{3},\frac{1}{3},\frac{1}{3})}$ phonon instability in
%   kagome metal ScV$_6$Sn$_6$} 

\author{Alaska Subedi} 

\affiliation{CPHT, CNRS, \'Ecole polytechnique, Institut Polytechnique
  de Paris, 91120 Palaiseau, France}

\date{\today}

\begin{abstract}
  Layered nickelates are believed to exhibit superconductivity similar
  to that found in the cuprates.  However, the precise crystal
  structure of the superconducting phase of the layered nickelates has
  not been fully clarified. Here, I use first principles calculations
  to study the pressure dependence of the structural instabilities in
  the single-layer-trilayer La$_3$Ni$_2$O$_7$, which is one member of
  the layered nickelates family that also shows signatures of
  superconductivity.  I find a nearly dispersionless nondegenerate
  phonon branch in the parent $P4/mmm$ phase that is unstable along
  the Brillouin zone edge $M$ $(\frac{1}{2}, \frac{1}{2}, 0)$
  $\rightarrow$ $A$ $(\frac{1}{2},\frac{1}{2},\frac{1}{2})$ at all
  investigated pressures up to 30 GPa.  Calculations show additional
  doubly-degenerate instabilities along the edge $MA$ at lower
  pressures. I used group-theoretical analysis to identify the
  distinct low-symmetry distortions possible due to these
  instabilities and generated them using the eigenvectors of the
  unstable modes. Structural relaxations show that the lowest energy
  structures at 0 and 10 GPa involve condensation of both the
  nondegenerate and doubly-degenerate instabilities, which is in
  contrast to the experimental refinements that involve condensation
  of only the doubly-degenerate branch.  I also find that
  structural distortions are energetically favorable at 20 GPa,
  contrary to the experiments that do not observe any distortions of
  the parent $P4/mmm$ structure at high pressures.
\end{abstract}

\maketitle

\section{Introduction}

Layered nickelates have been brought into sharp focus since signatures
of supercondutivity was reported in doped infinite-layer NdNiO$_2$
around 15 K in 2019 \cite{li2019}. This was soon followed by
observation of similar superconducting signatures in quintuple-layer
Nd$_6$Ni$_5$O$_{12}$ near 13 K \cite{Pan2021}.  Both findings were
made in thin-film samples at ambient pressure. More recently, the
exploration of layered nickelates under high pressure has yielded
remarkable results.  In 2023, La$_3$Ni$_2$O$_7$ was reported to
exhibit superconductivity at a notably high temperature of 80 K under
pressures exceeding 14 GPa \cite{Sun2023}. This high-pressure
superconductivity trend has extended to trilayer La$_4$Ni$_3$O$_{10}$
as well \cite{LiQ2024, Zhang2024, LiJ2024, ZhuY2024, Nagata2024}.

% This high-pressure superconductivity has also been observed in other
% layered nickelates, such as La4Ni3O10, under similar conditions
% \cite{LiQ2024, Zhang2024, LiJ2024, ZhuY2024, Nagata2024}.

%   This
% high-pressure superconductivity trend has extended to other layered
% nickelates as well, with reports of superconductivity in
% La$_4$Ni$_3$O$_{10}$ under similar conditions \cite{LiQ2024,
%   Zhang2024, LiJ2024, ZhuY2024, Nagata2024}.

Although substantial diamagnetic effects with large volume fraction
has been observed in infinite-layer Nd$_{0.8}$Sr$_{0.2}$NiO$_2$
\cite{Zeng2022} and La$_4$Ni$_3$O$_{10}$ \cite{Zhang2024,ZhuY2024},
only a weak diamagnetic response suggestive of filamentary or
inhomogenous superconductivity has been detected in La$_3$Ni$_2$O$_7$
\cite{Zhou2024,Wen2024}.  Nevertheless, La$_3$Ni$_2$O$_7$ exhibits
linear-$T$ resistivity above onset $T_c$ \cite{ZhangY2024}, motivating
similarity with the cuprates high-temperature superconductors.  This
has prompted extensive experimental and theoretical efforts to
elucidate the nature of superconductivity in this material
\cite{Hou2023,LiJ2024b,WangG2024,Ren2024,WangL2024,ZhangM2024,Liu2023,
  Khasanov2024,ChenX2024,ChenK2024,DanZ2024,LuoZ2023,ZhangY2023,Yang2023,
  Gu2023,Wu2024,Luo2024,Lu2024,Qu2024,Fan2024,Ouyang2024b,LiuY2023,Jiang2024,
  LiuH2023,YangY2023,ZhangY2024c,Christiansson2023,Labollita2024b,
  Labollita2024c,Lechermann2023,Lechermann2024,WangM2024}. However,
the crystal structure of the superconducting phase has not yet been
fully determined.

In this regard, it has recently been discovered that
La$_3$Ni$_2$O$_7$, in addition to the bilayer structure, also forms in
a single-layer trilayer (SL-TL) phase
\cite{Puphal2024,Chen2024,Wang2024}. Although no phase transitions
have been observed in thermodynamic experiments in this phase,
signatures of superconductivity have been found in resistivity
measurements at pressures above 8 GPa \cite{Puphal2024}.  X-ray
diffraction experiments find that the ambient-pressure structure of
the SL-TL phase can reasonably be refined to structures with $P4/mmm$,
$Fmmm$, $Cmmm$, or $Immma$ space groups
\cite{Puphal2024,Chen2024,Wang2024}, while the $P4/mmm$ structure well
describes the high-pressure phase above 12.8 GPa \cite{Puphal2024}.
Density functional theory calculations find that the $Fmmm$ and
$P4/mmm$ structures have the lowest energy for the SL-TL phase at 0
and 16 GPa, respectively \cite{ZhangY2024b}. Both experiment and
theory find that the electronic states near the Fermi level are
derived from nominally Ni $d_{z^2}$ narrow and $d_{x^2-y^2}$
dispersive bands
\cite{Puphal2024,Abadi2024,ZhangY2024b,LaBollita2024,Ouyang2024}.

\begin{figure*}
  \includegraphics[width=\textwidth]{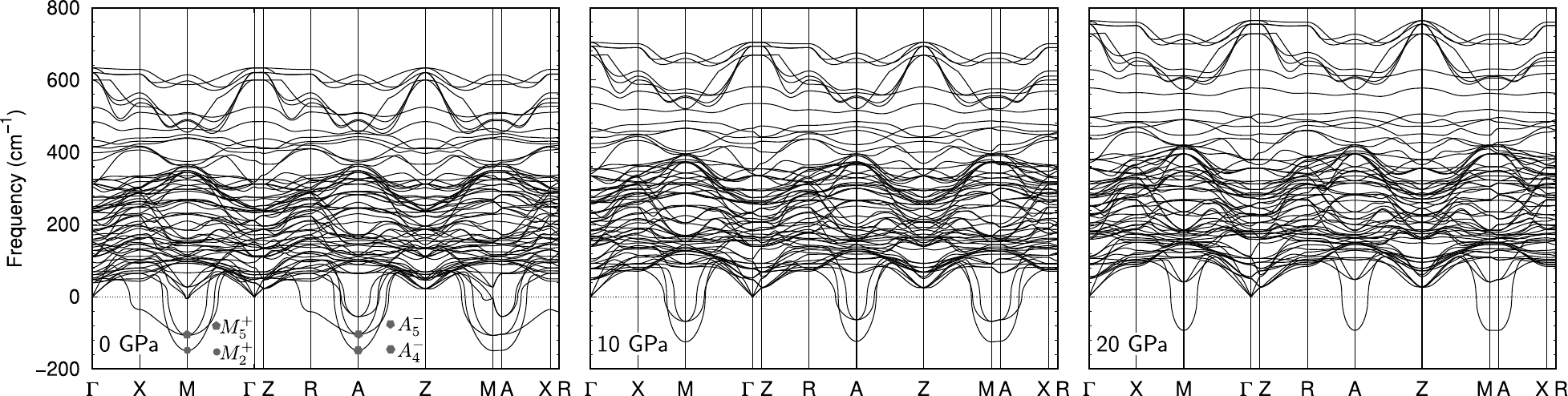}
  \caption{Calculated phonon dispersions of SL-TL La$_3$Ni$_2$O$_7$ at
    0, 10, and 20 GPa in the parent $P4/mmm$ phase.
    The high-symmetry points are $\Gamma$ $(0,0,0)$, $X$
    $(0,\frac{1}{2},0)$, $M$ $(\frac{1}{2}, \frac{1}{2}, 0)$, $Z$
    $(0,0,\frac{1}{2})$, $R$ $(0,\frac{1}{2},\frac{1}{2})$, and $A$
    $(\frac{1}{2},\frac{1}{2},\frac{1}{2})$ in terms of the reciprocal
    lattice vectors. Imaginary frequencies are indicated by negative
    values.}
  \label{fig:ph}
\end{figure*}

In this paper, I study the structural properties of SL-TL
La$_3$Ni$_2$O$_7$ as a function of pressure using first principles
calculations.  I consider the undistorted model of the SL-TL phase,
which has the space group $P4/mmm$, and calculate its phonon
dispersions.  The phonon dispersions at 0 GPa exhibit three unstable
branches along the Brillouin edge $M$ $(\frac{1}{2}, \frac{1}{2}, 0)$
$\rightarrow$ $A$ $(\frac{1}{2},\frac{1}{2},\frac{1}{2})$.  The most
unstable branch is nondegenerate, whereas the other two are doubly
degenerate.  As the pressure is increased to 10 GPa, the two lowest of
the three unstable branches remain unstable.  At 20 GPa, the
nondegenerate branch still remains unstable, while the
doubly-degenerate branches become stable.  I find that the $P4/mmm$
structure is unstable up to the maximum investigated pressure of 30
GPa, which is contrary to the experimental findings.  I used symmetry
analysis to enumerate possible distortions due to the two largest
instabilities at $M$ and $A$.  I generated these structures using the
eigenvectors of the unstable phonon modes and relaxed them to obtain
their total energies.  The lowest energy structures at 0 and 10 GPa
involve condensation of both the nondegenerate and doubly-degenerate
branches. This is at variance with the experimentally proposed $Fmmm$
and $Immma$ structures that involve condensation of only the
doubly-degenerate mode.  Additionally, the $Cmmm$ refinement cannot be
stabilized in the calculations.  At 0 GPa, I find that five distinct
lowest-energy structures lie within 0.4 meV/atom.  This near
degeneracy is increased at 10 GPa, with ten lowest-lying structures
having the same energy within numerical accuracy.  The degeneracy is
partially lifted and there are only three nearly degenerate low-energy
distortions at 20 GPa.  The large degeneracy of the distorted
structures is caused by the layered nature of SL-TL La$_3$Ni$_2$O$_7$
and the flatness of the unstable branches, which make the octahedral
rotations due to the phonon instabilities uncorrelated especially in
the out-of-plane direction.  This should manifest as a short coherence
length of the distortions in the experiments.

\section{Computational Approach}

The phonon dispersions presented in the paper were obtained using
density functional perturbation theory as implemented in the {\sc
  quantum espresso} package version 7.3.1 \cite{qe}.  This is a
pseudopotential-based package that uses plane-wave basis set.  I used
the version 1.0.0 of the ultrasoft pseudopotentials generated by Dal
Corso in my calculations \cite{pslib}.  They have the valence
configurations La $5s^2 5p^6 6s^{1.5} 5d^1 6p^{0.5}$, Ni $4s^2 3d^8$,
and O $2s^2 2p^4$. The calculations were performed within the
generalized gradient approximation of Perdew, Burke, and Ernzerhof
\cite{pbe}. The cutoffs of the basis-set and charge-density expansions
were set at 60 and 600 Ry, respectively. An $8 \times 8 \times 2$
$k$-point grid was used for the Brillouin zone integration with a
Marzari-Vanderbilt smearing of 0.01 Ry.  The dynamical matrices were
calculated on a $4 \times 4 \times 2$ grid, and the phonon dispersions
were obtained using Fourier interpolation.  I used the {\sc spglib}
\cite{spglib}, {\sc findsym} \cite{findsym}, and {\sc amplimodes}
\cite{ampli} packages in the symmetry analysis of the phonons and
distorted structures.  The {\sc isotropy} code \cite{isotropy} was
used to determine the order parameter directions due to the unstable
phonon modes, and these structures were generated on 192-atom $2
\times 2 \times 2$ supercells.

The structural relaxation of the distorted structures were performed
using the {\sc vasp} package version 6.4.2 for computational
efficiency \cite{vasp}. These calculations were performed using the
projector augmented-wave pseudopotentials with the valence
configurations of La $5s^2 5p^6 6s^2 4f^{0.0001} 5d^{0.9999}$, Ni
$3p^6 4s^1 3d^9$, and O $2s^2 2p^4$.  An $8 \times 8 \times 3$
$k$-point grid, 500 eV basis-set cutoff, and Methfesse-Paxton smearing
of 0.1 eV were used.  The spin-orbit interaction was neglected in all
calculations.

\begin{figure}%[ht]
  \includegraphics[width=0.98\columnwidth]{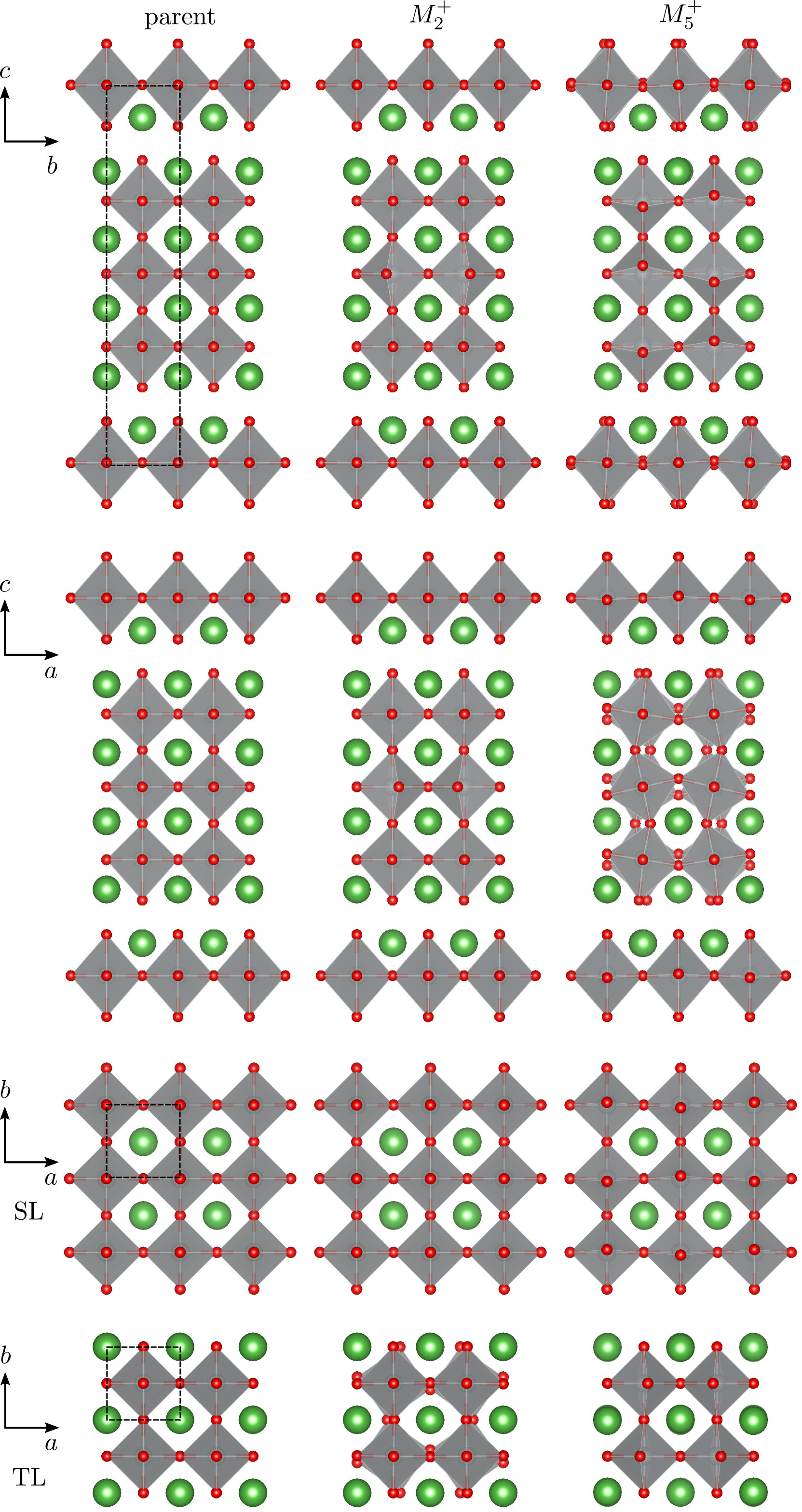}
  \caption{(Left) The undistorted $P4/mmm$ structure of SL-TL
    La$_3$Ni$_2$O$_7$ from different perspectives.  (Middle)
    Structural distortions due to the nondegenerate unstable phonon
    mode at $M$ that has the irrep $M_2^+$.  (Right) Structural
    distortiosns due to a component of the doubly-degenerate unstable
    mode at $M$ that has the irrep $M_5^+$.  The corresponding
    instabilities at $A$ involve the same atomic displacements but are
    out of phase in the adjacent unit cells along $c$.  The unit cell
    is indicated by the dashed black lines.  Green and red spheres
    indicate La and O, respectively.  Ni ions are inside the gray
    octahedra.}
  \label{fig:phdisp}
\end{figure}

\section{Results and Discussion}

The ambient pressure x-ray diffraction data of SL-TL \lno has been
refined using several different structural models.  Puphal \etal find
that an orthorhombic structure with the space group $Fmmm$ provides
the best refinement, but they also note that structures with the space
groups $Cmmm$, $P4/mmm$, and $Imma$ can provide satisfactory
refinements \cite{Puphal2024}.  Chen \etal report that the $Cmmm$
structure offers the best fit to their data, which can also be solved
using the $Imma$ structure \cite{Chen2024}. Wang \etal similarly
conclude that the $Cmmm$ structure provides the best fit to their data
\cite{Wang2024}.
The $Fmmm$, $Cmmm$, and $Imma$ structures of SL-TL \lno all derive
from the high-symmetry $P4/mmm$ structure via atomic displacements
along different unstable phonon coordinates.  To determine other
possible low-symmetry structures, I calculated the phonon dispersions
of this material at pressure values of 0, 10, and 20 GPa, which are
shown in Fig.~\ref{fig:ph}.  The calculated phonon dispersions exhibit
instabilities at all investigated pressures along the Brillouin zone
edge $MA$, which indicates the tetragonal $P4/mmm$ structure is
unstable at these pressures.

At 0 GPa, there are three unstable branches along $MA$.  The largest
instability occurs due to a nondegenerate branch that has the
irreducible representations (irreps) $M_2^+$ and $A_4^-$ at $M$ and
$A$, respectively.  The corresponding calculated imaginary frequencies
at $M$ and $A$ of this branch are 149$i$ and 148$i$ cm$^{-1}$.
This branch is followed above by a doubly degenerate branch with
calculated imaginary frequencies of 107$i$ and 104$i$ cm$^{-1}$ and
irreps of $M_5^+$ and $A_5^-$ at $M$ and $A$, respectively. Above this
lies another doubly degenerate branch with imaginary frequencies of
5$i$ and 55$i$ cm$^{-1}$ at $M$ and $A$, respectively.  As one can
see, the two most unstable branches are almost dispersionless along
$MA$, but the weakly unstable doubly-degenerate branch shows strong
dispersion.  Additionally, one offshoot of the more unstable doubly
degenerate branch is unstable throughout the face $MAXR$.

As the pressure is increased to 10 GPa, there are only two unstable
branches along $MA$.  The weakly-unstable doubly degenerate branch of
0 GPa becomes stable at 10 GPa.
%However, the two most unstable branches remain unstable along $MA$.
The calculated imaginary frequencies of the nondegenerate $M_2^+$ and
doubly-degenerate $M_5^+$ modes decrease to 125$i$ and 68$i$
cm$^{-1}$, whereas those of the nondegenerate $A_4^-$ and
doubly-degenerate $A_5^-$ modes decrease to 123$i$ and 64$i$
cm$^{-1}$, respectively.  It is noteworthy that the separation between
the two unstable branches increases relative to the values calculated
at 0 GPa.  For example, the difference between the two modes at $M$ is
42$i$ cm$^{-1}$ at 0 GPa, but it becomes 57$i$ cm$^{-1}$ at 10 GPa.

% The offshoot of the more unstable
% doubly degenerate branch also becomes stable along the edges $XR$,
% $XM$, and $RA$. 

At an even higher pressure of 20 GPa, only the nondegenerate branch is
unstable.  The calculated imaginary frequencies of the $M_2^+$ and
$A_4^-$ modes decrease to 93.0$i$ and 92.9$i$ cm$^{-1}$,
respectively. 
As the pressure is increased from 0 to 20 GPa, the frequencies of the
highest lying phonon branches that are dispersive and correspond to
inplane stretching motion of O ions shift up by more 130 cm$^{-1}$.
The less dispersive branches that lie just below them and correspond
to out-of-plane stretching motion of O ions increase their frequencies
by between 60--100 cm$^{-1}$.  This smaller increase is not surprising
considering the weaker bond in out-of-plane direction due to the
presence interlayer spacing.  The frequencies of the rest of the
stable branches increase by less then 50 cm$^{-1}$.
Incidentally, I did calculations also at 30 GPa and found that this
branch remains unstable, which is surprising considering that Puphal
\etal report that the structure above $\sim$12 GPa has the space group
$P4/mmm$ \cite{Puphal2024}.

The left column of Fig.~\ref{fig:phdisp} shows the parent $P4/mmm$
structure along different directions.  The middle column shows the
atomic displacements due to the most unstable nondegenerate branch at
$M$ that has the irrep $M_2^+$ and 0 GPa frequency of 149$i$
cm$^{-1}$.  This mode mainly involves inplane rotation of the middle
layer of the NiO$_6$ octahedra within the trilayer.  The NiO$_6$
octahedra in the outer layers sandwiching the middle layer also show
weak inplane rotations out-of-phase along $c$ that is difficult to
discern in the figure.  However, the NiO$_6$ octahedra within the
single layer and La and Ni ions remain fixed.  The displacement
pattern due to the corresponding mode of this branch at $A$ is
similar, except that they are out of phase by 180$^{\circ}$ in the
neighboring unit cells in the out-of-plane direction.

The right column of Fig.~\ref{fig:phdisp} shows the atomic
displacements due to one component of the second-most unstable phonon
mode at $M$ at 0 GPa that has the irrep $M_5^+$ and 0 GPa frequency of
107$i$ cm$^{-1}$.  This mode involves rotation of all the NiO$_6$
octahedra in planes parallel to the $c$ axis.  For the component shown
in the figure, the NiO$_6$ octahedra within the trilayer rotate in the
$ac$ plane, and these rotations are out-of-phase along the $b$ axis.
Meanwhile, the octahedra within the single layer rotate in the $bc$
plane, and they propagate out-of-phase along the $a$ axis.  The
rotations of the octahedra within single layer are weaker than those
exhibited by the octahedra within the trilayer.  Additionally, the
outer Ni ions within the trilayer and all the La ions also move by
small amounts that are difficult to notice in the figure.  The Ni ions
and the La ions within the single layer displace along the $a$ axis,
while the La ions within the trilayer displace along the $b$ axis.
These displacements propagate out-of-phase along all three directions.
% The displacements due to the corresponding mode at $A$ has the same
% pattern other than being out-of-phase in the neighboring unit cells.

These unstable phonons of SL-TL La$_3$Ni$_2$O$_7$ generate many
possible low symmetry structures.  To systematically search for the
lowest-energy structure, I used the {\sc isotropy} package to
enumerate the symmetrically distinct structural distortions due to
these instabilities.  I considered the two lowest-frequency modes at
$M$ and $A$ that have the irreps $M_2^+$, $M_5^+$, $A_4^-$, and
$A_5^-$, as well as their direct sum with each other. I found 26
distinct order parameters.  I generated all these structures using the
eigenvectors of the unstable phonon modes and fully relaxed them at
specific pressures.

\begin{table}
    \caption{\label{tab:zero} Isotropy subgroups and corresponding
      order parameters of $P4/mmm$ due to the irreps $M_2^+$, $M_5^+$,
      $A_4^-$, and $A_5^-$, and their direct sum with each other.
      Total energies of the structures corresponding to these order
      parameters directions after structural relaxations at 0 GPa are
      given in the units of meV/atom relative to the parent $P4/mmm$
      phase.  Not all distortions could be stabilized. Structures with
      '' in the energy column relaxed to the preceding order
      parameter. }

    \begin{ruledtabular}
      
      \begin{tabular}{l l l l l d{3.2}}
        % \begin{tabular}{l l c}
        % Space group (No.)  & OPD & Energy (meV/\fu) \\
        Space group     & $M_2^+$ & $A_4^-$ & $M_5^+$ & $A_5^-$  & \multicolumn{1}{c}{Energy} \\
        \hline
        $P4/mmm$        &         &         &         &          &  0.0 \\
        $P4/mbm$        & $(a)$   & $(b)$   &         &          & -1.9 \\
        $Cmme$          &         &         & $(a,a)$ & $(b,-b)$ & -3.2 \\
        $Fmmm$          &         &         &         & $(a,a)$  & -3.2 \\
        $Pmna$          &         &         & $(a,0)$ & $(b,0)$  & -3.6 \\
        $Imma$          &         &         &         & $(a,0)$  & -3.6 \\
        $C2/m$          &         &         &         & $(a,b)$  & \multicolumn{1}{c}{''} \\ % -3.61 Imma
        $P4/mbm$        & $(a)$   &         &         &          & -3.9 \\
        $I4/mcm$        &         & $(a)$   &         &          & -3.9 \\
        $Cmme$          &         &         & $(a,a)$ &          & -4.7 \\
        $Ccce$          &         &         & $(a,a)$ & $(b,b)$  & -4.7 \\
        $P2/c$          &         &         & $(a,b)$ & $(c,d)$  & \multicolumn{1}{c}{''} \\ % -4.69 Ccce
        $Cmcm$          & $(a)$   &         &         & $(b,b)$  & -5.1 \\
        $C2/m$          &         & $(a)$   &         & $(b,b)$  & -5.1 \\
        $Pmna$          &         &         & $(a,0)$ &          & -5.2 \\
        $Pnna$          &         &         & $(a,0)$ & $(0,-b)$ & \multicolumn{1}{c}{''} \\ % -5.19 Pmna
        $P2/c$          &         &         & $(a,b)$ &          & \multicolumn{1}{c}{''} \\ % -5.21 Pmna
        $Pnma$          & $(a)$   &         &         & $(b,0)$  & -5.4 \\
        $P2_1/m$        & $(a)$   &         &         & $(b,c)$  & \multicolumn{1}{c}{''} \\ % -5.40 Pnma
        $C2/c$          &         & $(a)$   &         & $(b,0)$  & -5.4 \\
        $P\overline{1}$ &         & $(a)$   &         & $(b,c)$  & \multicolumn{1}{c}{''} \\ % -5.40 C2/c
        $C2/m$          & $(a)$ &       & $(b,b)$ &  & -6.3 \\
        $Cmce$          &       & $(a)$ & $(b,b)$ &  & -6.3 \\
        $P2_1/c$        &       & $(a)$ & $(b,c)$ &  & -6.5 \\
        $Pbcn$          &       & $(a)$ & $(b,0)$ &  & -6.7 \\
        $P2_1/c$        & $(a)$ &       & $(b,0)$ &  & -6.7 \\
        $P\overline{1}$ & $(a)$ &       & $(b,c)$ &  & \multicolumn{1}{c}{''} \\ % -6.7 P2_1/c
      \end{tabular}
    \end{ruledtabular}
\end{table}

% At 0 GPa:
% show the isotropy subgroups
% discuss the energy rankings

Table~\ref{tab:zero} shows the calculated energies of the relaxed
structures at 0 GPa.  One can first note that the $Cmmm$ phase refined
by Chen \etal and Wang \etal does not appear in the table
\cite{Chen2024,Wang2024}.  When I took their experimental structures
and performed structural relaxation calculations, they relaxed to the
$P4/mmm$ phase.  This suggests that the $Cmmm$ phase is not
energetically favorable.  In any case, I find that nineteen distinct
distorted structures due to the phonon instabilities could be
stabilized that have energies lower than that of the parent $P4/mmm$
phase.  The distorted structures $A_4^-(a) + M_5^+(b,0)$ with the
space group $Pbcn$ and $M_2^+(a) + M_5^+(b,0)$ with the space group
$P2_1/c$ have the lowest energy, with a value of $-6.7$ meV/atom
relative to the undistorted $P4/mmm$ phase.  These structures involves
condensation of the lowest-lying nondegenerate branch at $A$ and $M$,
respectively, along with the $M_5^+(b,0)$ order parameter direction of
the doubly-degenerate unstable mode at $M$.
The $Imma$ and $Fmmm$ structures suggested by the experiments, with
the respective order parameters of $A_5^-(a,0)$ and $A_5^-(a,a)$,
derive from condensation of the doubly-degenerate $A_5^-$ instability.
These structures show smaller energy gains of $-3.6$ and $-3.2$
meV/atom, respectively.
Interestingly, the $P4/mbm$ phase with the $M_2^+(a)+A_4^-(b)$
distortion that arises due to condensation of the largest
instabilities at $M$ and $A$ leads to the smallest energy gain.
The largest energy gains involves condensation of the $M_5^+$
instability in combination with either $M_2^+$ or $A_4^-$
instabilities, and there are five of them within 0.4 meV/atom of each
other.  These structures mainly differ in how the NiO$_6$ octahedral
rotations due to the phonon instabilities propagate along the three axes.

The five lowest energy structures are not the only group of nearly
degenerate structures in Table~\ref{tab:zero}.  The table shows that
condensation of either $M_2^+$ or $A_4^-$ along with the
doubly-degenerate $A_5^-$ mode also leads to almost the same energy.
This derives from the layered structure of SL-TL La$_3$Ni$_2$O$_7$ and
the fact that the unstable branch encompassing both $M_2^+$ and $A_4^-$ is
almost dispersionless.  Since the NiO$_6$ octahedra of the SL are not
connected to those of the TL along the $c$ axis, their rotations can
be uncorrelated along $c$.  So the condensation of the dispersionless
unstable branch at any out-of-plane periodicity will be nearly
degenerate.  The close energies of these structures should result in
especially short coherence length along the out-of-plane direction,
making it difficult to resolve the sample to a single phase in the
experiments.

\begin{table}
  \caption{\label{tab:ten} Total energies of the structures
    corresponding to the $M_2^+$, $A_4^-$, $M_5^+$, and $A_5^-$
    instabilities after structural relaxations at 10 GPa in the units
    of meV/atom relative to the parent $P4/mmm$ phase.  Not all
    distortions could be stabilized.  Structures with '' in the energy
    column relaxed to the preceding order parameter.}
  \begin{ruledtabular}
    \begin{tabular}{l l l l l d{3.2}}
      % \begin{tabular}{l l c}
      % Space group (No.)  & OPD & Energy (meV/\fu) \\
      Space group     & $M_2^+$ & $A_4^-$ & $M_5^+$ & $A_5^-$  & \multicolumn{1}{c}{Energy} \\
      \hline
      $P4/mmm$        &         &         &         &          &  0.0 \\
      $Cmme$          &         &         & $(a,a)$ & $(b,-b)$ & -0.2 \\
      $Pmna$          &         &         & $(a,0)$ & $(b,0)$  & -0.3 \\
      $Fmmm$          &         &         &         & $(a,a)$  & -0.4 \\
      $Imma$          &         &         &         & $(a,0)$  & -0.4 \\
      $C2/m$          &         &         &         & $(a,b)$  & -0.5 \\
      $Pnna$          &         &         & $(a,0)$ & $(0,-b)$ & -0.5 \\ 
      $Ccce$          &         &         & $(a,a)$ & $(b,b)$  & -0.5 \\
      $P2/c$          &         &         & $(a,b)$ & $(c,d)$  & \multicolumn{1}{c}{''} \\ % -0.38 Ccce
      $Cmme$          &         &         & $(a,a)$ &          & -0.6 \\
      $P2/c$          &         &         & $(a,b)$ &          & \multicolumn{1}{c}{''} \\ % -0.53 Cmme
      $Pmna$          &         &         & $(a,0)$ &          & -0.7 \\
      $P4/mbm$        & $(a)$   & $(b)$   &         &          & -0.9 \\
      $P4/mbm$        & $(a)$   &         &         &          & -1.9 \\
      $I4/mcm$        &         & $(a)$   &         &          & -1.9 \\
      $Cmcm$          & $(a)$   &         &         & $(b,b)$  & -1.9 \\
      $P2_1/m$        & $(a)$   &         &         & $(b,c)$  & \multicolumn{1}{c}{''} \\ % -1.45 Cmcm
      $Pnma$          & $(a)$   &         &         & $(b,0)$  & -1.9 \\
      $Cmce$          &         & $(a)$   & $(b,b)$ &          & -1.9 \\
      $P2_1/c$        &         & $(a)$   & $(b,c)$ &          & \multicolumn{1}{c}{''} \\ % -1.50 Cmce
      $Pbcn$          &         & $(a)$   & $(b,0)$ &          & -1.9 \\
      $C2/m$          & $(a)$   &         & $(b,b)$ &          & -1.9 \\
      $P\overline{1}$ & $(a)$   &         & $(b,c)$ &          & \multicolumn{1}{c}{''} \\ % -1.48 C2/m
      $P2_1/c$        & $(a)$   &         & $(b,0)$ &          & -1.9 \\
      $C2/m$          &         & $(a)$   &         & $(b,b)$  & -1.9 \\
      $P\overline{1}$ &         & $(a)$   &         & $(b,c)$  & \multicolumn{1}{c}{''} \\ % -1.47 C2/m
      $C2/c$          &         & $(a)$   &         & $(b,0)$  & -1.9 \\
      \end{tabular}
    \end{ruledtabular}
\end{table}

% 10 GPa
The results of structural relaxations at 10 GPa are shown in
Table~\ref{tab:ten}.  As expected from the instabilities becoming
weaker as the pressure is increased, the energy gains due to the
distortions at 10 GPa are smaller than those at 0 GPa.  Another
difference is the larger number---ten---of structures having the
lowest energy gain of $-1.9$ eV/atom.  These structures involve
condensation of the nondegenerate mode by itself and also in
combination with the doubly-degenerate mode.  
% These contain condensation of
% the doubly-degenerate mode at either $M$ or $A$ along with the
% nondegenerate mode at either $M$ or $A$.  
In contrast, at 0 GPa,
structures involving condensation of the doubly-degenerate mode at $A$
and the nondegenerate branch were at least 0.9 meV/atom higher than
the structures involving condensation of the doubly-degenerate mode at
$M$ and the nondegenerate branch.  
%
% Like the
% nearly-degenerate lowest energy structures at 0 GPa, these structures
% are mainly distinguished by the different propagations of the
% octahedral rotation along the three axis.

There are further minor differences in the energetics of the
distortions at 0 and 10 GPa even though they involve the same two
unstable phonon branches.  The $Pnna$ structure with the order
parameter $M_5^+(a,0)+A_5^-(0,-b)$ can additionally be stabilized at
10 GPa.  Furthermore, the energy gains at 10 GPa due to the
condensation of only the doubly-degenerate branch is smaller than
those due to the condensation of only the nondegenerate branch. At 0
GPa, two structures due to the condensation of the doubly-degenerate
branch are lower in energy than the structures that contain only the
distortions due to the nondegenerate branch. The larger separation
between the two unstable phonon branches at 10 GPa likely explains the
change in the ranking of the relative energies.  It is also noteworthy
that the additional energy gain due to the condensation of the
doubly-degenerate branch on top of the distortions due to the
nondegenerate branch is negligible at 10 GPa, which may also be caused
by the higher relative position of the doubly-degenerate branch.

% Although the two phonon branches remain unstable along $MA$ as the
% pressure is increased to 10 GPa, the ranking of the relative energies
% of the isotropy subgroups after structural relaxations are different
% from that obtained for 0 GPa, as one can see in Table II.  Another
% difference is that only NN distinct distortions could be stabilized  % TODO
% after relaxations at 10 GPa, in contrast to the 21 obtained at 0 GPa.
% Furthermore, the energy gains due to the distortions are also smaller,
% which is consistent with the instabilities becoming weaker as the
% pressure is increased.  At 10 GPa, the $C2/m$ structure with the order
% parameter $M_2^+(a) + M_5^+(b,b)$ has the lowest energy, with an
% energy gain of $-2.5$ meV/atom relative to the parent $P4/mmm$ phase
% at this pressure. Interestingly, the energy gain due to the
% condensation of only the nondegenerate most unstable branch is $-2.4$
% meV/atom. The smaller additional energy gain due to the combined
% condensation of the two unstable branches likely reflects the fact
% that the difference in the imaginary frequencies of these branches are
% bigger at 10 GPa.  
% In any case, the MM lowest-energy distinct structures are
% only within 0.1 eV of each other, and this should make it highly
% challenging to refine experimental sample to a single phase.

\begin{table}
  \caption{\label{tab:twenty} Total energies of the structures
    corresponding to the $M_2^+$ and $A_4^-$ instabilities after
    structural relaxations at 20 GPa in the units of meV/atom relative
    to the parent $P4/mmm$ phase.  The structures corresponding to the
    $M_5^+$ and $A_5^-$ instabilities relaxed to the parent phase
    because these modes are stable at this pressure.}
  \begin{ruledtabular}
    \begin{tabular}{l l l d{3.2}}
      % \begin{tabular}{l l c}
      % Space group (No.)  & OPD & Energy (meV/\fu) \\
      Space group     & $M_2^+$ & $A_4^-$  & \multicolumn{1}{c}{Energy} \\
      \hline
      $P4/mmm$        &         &          &  0.0 \\
      $P4/mbm$        & $(a)$   & $(b)$    & -0.2 \\
      $P4/mbm$        & $(a)$   &          & -0.4 \\
      $I4/mcm$        &         & $(a)$    & -0.4 \\
      \end{tabular}
    \end{ruledtabular}
\end{table}

% % 20 GPa

At 20 GPa, only the $M_2^+(a)$, $A_4^-(a)$, and $M_2^+(a) + A_4^-(a)$
order parameters with the respective space groups of $P4/mbm$,
$I4/mcm$, and $P4/mbm$ are stable after structural relaxations. This
is as expected since only the nondegenerate branch is unstable at this
pressure.  The total energies of these structures are given in
Table~\ref{tab:twenty}.  The three distorted structures lie within 0.2
meV/atom of each other, which again reflects the nearly dispersionless
nature of the unstable branch along $MA$.  The $M_2^+(a)$ and
$A_4^-(a)$ distortions have the lowest energy of $-0.4$ meV/atom
relative to the parent $P4/mmm$ phase.  Despite the small energy gain,
the oxygen octahedra in the middle layer of the trilayer rotate by
16.5$^\circ$ in these structures.  However, this layer is sandwiched
by the outer layers that rotate by less than 3$^{\circ}$, and this
probably makes it difficult to resolve this structure in the
experiments.

%% TODO: electronic structure

\section{Summary and Conclusions}

In summary, I used first principles phonon and structural relaxation
calculations to show that the undistorted $P4/mmm$ model of SL-TL
La$_3$Ni$_2$O$_7$ is unstable at all investigated pressures up to 30
GPa.  Calculated phonon dispersions show an unstable nondegenerate
branch at all pressures along the Brillouin zone edge $MA$. At 0 and
10 GPa, there are additionally two and one doubly-degenerate unstable
branches, respectively, along this path.  I used group-theoretical
analysis to find the possible structural distortions that could arise
due to the two most unstable phonon modes at $M$ and $A$. Structures
corresponding to these order parameter directions were constructed
using the eigenvectors of the unstable phonon modes.  They were then
relaxed at specific pressures to obtain their total energies.

I find that the lowest-energy structures at 0 and 10 GPa involve
condensation of both the nondegenerate and doubly-degenerate unstable
phonon branches.  The $Fmmm$ and $Imma$ experimental refinements,
which involve the condensation of only the doubly-degenerate mode at
$A$, lie higher in energy.  Another experimentally proposed structure
$Cmmm$ could not be stabilized.  The energy gain due to the
distortions are relatively small, consistent with the small values of
the imaginary frequencies of the unstable phonon modes.  At 0 GPa, the
lowest-energy structures show an energy gain of $-6.7$ meV/atom
relative to the parent phase.  Furthermore there are five distinct
structures within 0.4 meV/atom of this value.  This degeneracy
increases at 10 GPa, where there are ten distinct structures showing
the largest energy gain of $-1.9$ meV/atom.  Only the nondegenerate
branch is unstable at 20 GPa.  This reduces the degeneracy, with the
three distorted structures that are energetically lower than the
parent phase lying 0.2 meV/atom of each other.  The structural
degeneracies occur due to the layered nature of this material and the
flatness of the unstable phonon branches, which combine to make the
octahedral rotations associated with the phonon instabilities
especially uncorrelated in the out-of-plane direction.  Furthermore,
the nearly dispersionless phonon instabilities could condense at any
intermediate point in the Brillouin zone edge.  This should manifest
as short coherence length of the out-of-plane structural distortions
in the experiments.

\section{acknowledgements}
This work was supported by GENCI-TGCC under grant no.\ A0170913028.

\bibliography{la3ni2o7-1313-paper}

\end{document}